# Improved Gate Reliability of *p*-GaN Gate HEMTs by Gate Doping Engineering

Guangnan Zhou, Fanming Zeng, Rongyu Gao, Qing Wang, Kai Cheng, Guangrui Xia, Hongyu Yu

*Abstract*—We present a novel *p*-GaN gate HEMT structure with reduced hole concentration near the Schottky interface by doping engineering in MOCVD, which aims at lowering the electric field across the gate. By employing an additional unintentionally doped GaN layer, the gate leakage current is suppressed and the gate breakdown voltage is boosted from 10.6 to 14.6 V with negligible influence on the threshold voltage and on-resistance. Time-dependent gate breakdown measurements reveal that the maximum gate drive voltage increases from 6.2 to 10.6 V for a 10-year lifetime with a 1% gate failure rate. This method effectively expands the operating voltage margin of the *p*-GaN gate HEMTs without any other additional process steps.

## I. INTRODUCTION

Gallium nitride (GaN) power transistors have drawn increasing attention in high-power, high-efficiency power conversion systems owing to their high breakdown voltage, fast switching speed and low ON-resistance [1-3]. In practical applications, enhancement-mode (E-mode) HEMTs with the normally-off operation are preferred for safety considerations. Among different approaches to realize E-mode HEMTs [4-7], the *p*-GaN gate AlGaN/GaN HEMT emerged as a leading solution [8, 9].

However, due to the relatively low gate breakdown voltage (BV) (usually 10 - 12 V), the maximum gate operation voltages ($V_{G-max}$) for *p*-GaN gate HEMTs are usually between 6 - 8 V [8, 9]. The small gate voltage swings have imposed significant constraints on the gate driver design and resulted in lifetime reduction. It is highly desired to increase the gate BV and the $V_{G-max}$ for a wider gate drive window.

Several reliability studies focused on the *p*-GaN gate HEMTs have been carried out over the last few years [10-24]. Different failure mechanisms have been proposed, including: i) metal/*p*-GaN Schottky junction breakdown, which is widely accepted and most likely to happen prior to others [10-14]; ii) the passivation/*p*-GaN sidewall related breakdown, which is a bigger problem in self-aligned gate metal/*p*-GaN structure [15, 16]; and iii) *p*-GaN/AlGaN/GaN junction breakdown [17, 18]. Many researchers have been tackling the gate breakdown challenge with various methods [19-23]. Zhou *et al.* [19] and Zhang *et al.* [20] adopted special treatments in the gate-stacks to enhance the *p*-GaN/metal Schottky junction, where suppressed gate leakage currents and enhanced gate BV were achieved. An *n*-GaN/*p*-GaN/AlGaN/GaN epitaxial structure was proposed for gate reliability enhancement [22], where the Schottky junction was replaced by a metal-n-p junction. However, a high-temperature activation of *p*-GaN after the gate etching is required, which can be detrimental to the channel due to the strain relaxation of AlGaN [25, 26]. Thus, a technique based on the commonly used *p*-GaN gate HEMT process would be favored owing to the process simplicity.

In this work, we demonstrated that using an additional unintentionally-doped GaN (*u*-GaN) layer on top of *p*-GaN could effectively lower the electric field in the gate and thus enhance the reliability of HEMTs without additional annealing step. This *u*-GaN/*p*-GaN gate method neither impairs other electrical characteristics nor requires any extra process steps.

## II. DEVICE STRUCTURE AND FABRICATION

The *p*-GaN gate HEMTs were fabricated on 85 nm *p*-GaN/15-nm Al$_{0.2}$Ga$_{0.8}$N/0.7-nm AlN/4.5-µm GaN epitaxial structure grown by metal-organic chemical vapor deposition (MOCVD) on 2-inch Si (111) substrates from Enkris Semiconductor Inc, as shown in Fig. 1(a). The *p*-GaN layer was doped with Mg to a concentration of $4 \times 10^{19}$ cm$^{-3}$. Three device structure types, named structure A, B and C, have been adopted, as illustrated in Fig. 2(b). They have identical epi-structure and growth conditions except that structure B and structure C have additional epitaxial *u*-GaN layers on top of the *p*-GaN grown by MOCVD, which was used to reduce the hole concentration in the region close to the Schottky interface with

This work was supported by Grant #2019B010128001 and #2019B010142001 from Guangdong Science and Technology Department, Grant #61704004 from Natural Science Foundation of China, Grant #JCYJ20180305180619573 and #JCYJ20170412153356899 from Shenzhen Municipal Council of Science and Innovation. (Corresponding authors: G. Xia and H. Yu.)

G. Zhou is with and the School of Microelectronics, Southern University of Science and Technology (SUSTech), and also with the Department of Materials Engineering, the University of British Columbia (UBC), Vancouver, BC V6T 1Z4, Canada.
F. Zeng, R. Gao, Q. Wang, and H. Yu are with the School of Microelectronics, SUSTech; Engineering Research Center of Integrated Circuits for Next-Generation Communications, Ministry of Education; Shenzhen Institute of Wide-bandgap Semiconductors; and GaN Device Engineering Technology Research Center of Guangdong, SUSTech, 518055 Shenzhen, Guangdong, China. (e-mail: yuhy@sustech.edu.cn)
K. Cheng is with the Enkris Semiconductor Inc. Suzhou, 215123, China.
G. Xia is with the Department of Materials Engineering, UBC, Vancouver, BC V6T 1Z4, Canada. (e-mail: gxia@mail.ubc.ca)

the metal, thus widening the depletion region and reducing the maximum electric field in the GaN gate. The thicknesses of *u*-GaN were 20 and 30 nm for structure B and C, respectively. Due to the memory effect and diffusion of Mg, the *u*-GaN layers should be slightly *p*-type doping after processing [27].

The fabrication process started with a gate definition by a Cl-based plasma etch. Devices with structure A, B and C have an identical fabrication process except that B and C have longer GaN etching times. Thanks to the high-selectivity and low etching rate of the etching recipe, a smooth AlGaN surface with low etching damages has been obtained for all three structures (Fig. 2(b)). After the gate contact window opening, a Schottky-type contact was formed between the Ti/Au and the GaN gate. The devices tested feature a gate width ($W_G$) of 100 μm, a gate length ($L_G$) of 5 μm, a gate-source distance ($L_{GS}$) of 5 μm, and a gate-drain distance ($L_{GD}$) of 15 μm. The relatively longer $L_{GS}$ compared to literature is to suppress the gate breakdown induced by the passivation/*p*-GaN sidewall [12, 15]. On-wafer characterizations were performed by a Keithley 4200 analyzer.

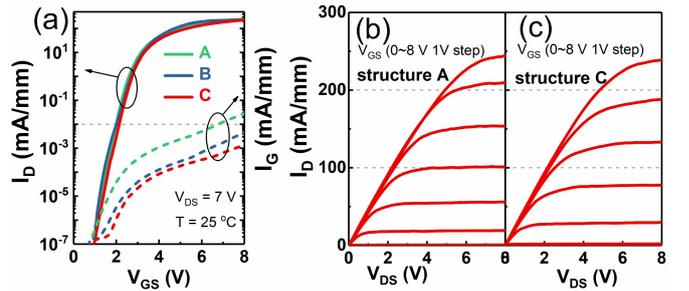

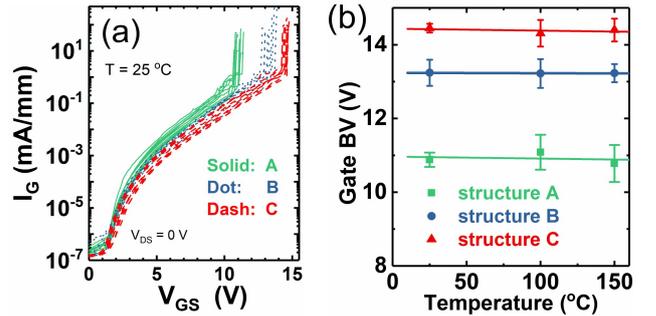

Fig. 2. (a) Transfer characteristic of the devices with structure A, B, and C; (b) output characteristic of A; (c) output characteristic of C.

Fig. 3. (a) Gate leakage and breakdown characteristics of the devices with structure A, B and C at 25 °C, and;(b) at 150 °C; (c) the BV of different structures and their temperature dependence.

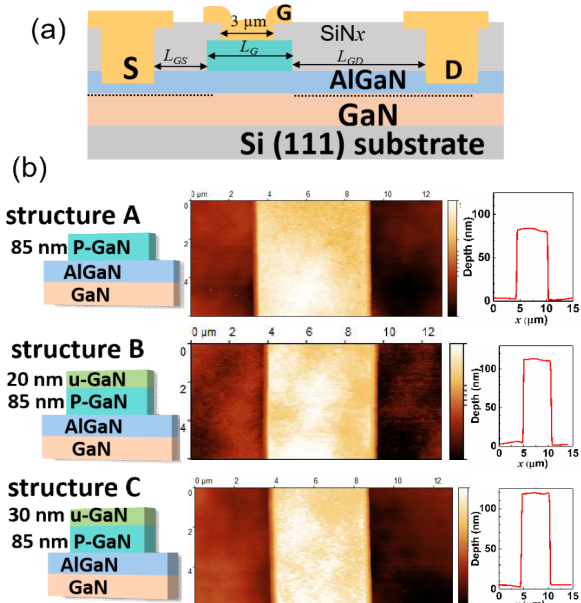

Fig.1. (a) Cross-sectional schematic of the *p*-GaN gate HEMTs; (b) Schematic of the gate structure A, B, C after the gate definition by *u*-GaN/*p*-GaN etch (left column) and their corresponding AFM characterization results (middle column), and their extracted profiles (right column).

## III. DEVICE CHARACTERIZATIONS
### A. Static Performance and Temperature Dependence

Fig. 2(a) depicts the transfer characteristics of the devices with structure A, B, and C. All devices exhibit a $V_{TH}$ of ~ 2.1 V (defined at $I_D$ = 0.01 mA/mm), and an ON/OFF ratio larger than $10^8$. Fig. 2(b) and (c) show the output characteristic of devices with structure A and C, in both of which a $R_{ON}$ of 22 Ω·mm is obtained. These results show that the doping engineering has a negligible effect on the conduction characteristics (e.g., $R_{ON}$ and $V_{TH}$), consistent with [12, 20]. Meanwhile, a clear impact can be observed on the gate leakage characteristics as shown in Fig. 2(a) and Fig. 3(a). Under a forward bias, the gate leakage current is dominant by the metal/GaN Schottky junction [12]. Owing to the reduced acceptor doping concentration in *u*-GaN, structure B and C show lower gate leakage under forward bias.

Fig. 3(a) compares the forward gate breakdown characteristics of A, B and C measured at 25 °C. The breakdown mechanism was determined to be Schottky junction failure [14]. Compared to A, the additional *u*-GaN layer in B has effectively boosted the BV from 10.9 V to 13.2 V. For C with a thicker *u*-GaN layer, the gate BV further reached 14.6 V. By introducing the *u*-GaN layer between the *p*-GaN and metal, the hole concentration close to the Schottky junction can be effectively decreased, which promotes a wider depletion region when a positive bias is applied. Thus, the peak electric field is lowered and the gate BV is significantly enlarged. Tallarico *et al.* have adopted a similar strategy by Mg doping compensation in [12]. However, no significant gate BV increase has been observed in their work. This difference can be attributed to their short $L_{GS}$ and self-aligned metal/*p*-GaN architecture, which makes the *p*-GaN sidewall vulnerable [14-16].

Fig. 3(b) shows the statistical summary of the gate BVs of structure A, B and C at 25 °C, 100 °C and 150 °C. For each temperature, at least fifteen devices were measured for each structure. All the gate BVs of the three structures show weak dependences on the temperature. These results demonstrate that the doping engineering can effectively suppress the gate leakage current and boost the gate BV without impacting the threshold voltage or any additional fabrication process.

### B. TDGB Analysis

Time-dependent gate breakdown (TDGB) tests with constant voltage stresses were performed to evaluate the gate reliability of the samples with structure A (Fig. 4) and C (Fig. 5). A constant voltage is applied on the gate with $V_{DS}$ = 0 V at room temperature. The time-to-breakdown ($t_{BD}$) is defined as when the gate leakage shows a sudden increase. For each structure, three different $V_{GS}$ were adopted (10 V, 10.2 V, 10.4 V for structure A, and 12.8 V, 13 V, 13.2 V for structure C). $t_{BD}$ distribution can be described by the Weibull statistics. The shape factor $\beta$ extracted from structure A is 0.74 - 0.82, whereas it is 0.76 - 0.87 from structure C. The comparable $\beta$ values indicate a similar degradation mechanism.

The lifetime prediction was performed using the most conservative exponential law (i.e., linear fitting of the $\ln(t_{BD})$-$V_{GS}$ relationship), as shown in Fig. 4(b) and Fig. 5(b). Considering a 10-year lifetime at a failure level of 1%, the $V_{G-max}$ is determined to be 6.2 V for structure A. Meanwhile, a much higher $V_{G-max}$ of 10.6 V has been achieved in structure C. The increased applicable gate voltage range offers more gate driver design flexibility and robust gate reliability.

Fig. 6 plots $V_{G-max}$ and the corresponding $V_{TH}$ observed in the p-GaN gate HEMTs in this work and other p-GaN gate HEMTs in literature [10, 15-17, 22, 24]. A high $V_{TH}$ and large $V_{G-max}$ are particularly desired in p-GaN gate HEMTs. The $V_{G-max}$ values are extracted by the exponential law for a 10-year lifetime at a failure rate of either 1% or 63%. Structure C demonstrates the largest $V_{G-max}$.

Moreover, this technique requires no additional process step after the MOCVD growth. The proposed u-GaN/p-GaN structure is highly effective in improving the p-GaN gate HEMT reliability for high-efficiency power conversion systems.

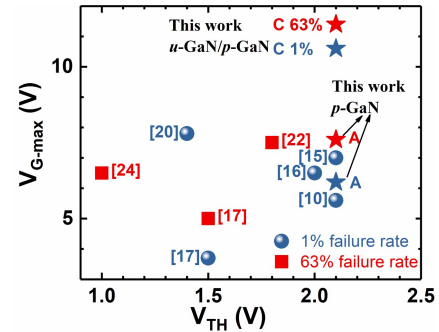

Fig. 6 Comparison of $V_{G-max}$ and $V_{TH}$ of structure A and C and other p-GaN gate HEMTs. The blue ones are estimated at a failure rate of 1%, and the red ones are estimated at 63%.

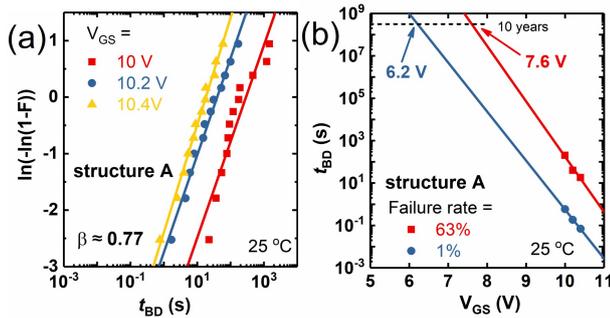

Fig. 4. (a) Weibull plot of $t_{BD}$ distribution of structure A and; (b) Lifetime prediction. By choosing a 63% and 1% failure rate for a 10-year lifetime, the maximum applicable $V_{GS}$ are 6.2 V and 7.6 V respectively.

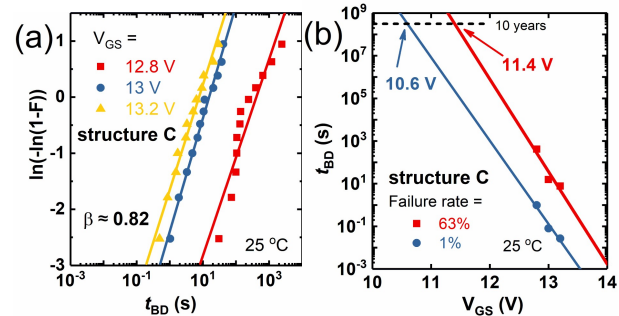

Fig. 5. (a) Weibull plot of $t_{BD}$ distribution of structure C and; (b) Lifetime prediction. By choosing a 63% and 1% failure rate for a 10-year lifetime, the maximum applicable $V_{GS}$ are 11.4 V and 10.6 V respectively.

## IV. CONCLUSION

In this work, a novel p-GaN gate HEMT structure with doping engineering was proposed and investigated. By growing an additional u-GaN layer on top of the p-GaN, more robust and reliable devices with lower gate leakage currents, higher gate BV (14.6 V), larger $V_{G-max}$ (10.6 V) were obtained, with negligible impacts on the the $V_{TH}$, $R_{ON}$ and sub-threshold slopes.